\documentclass[aps,prl,groupedaddress,showpacs]{revtex4}
\usepackage{graphicx}
\usepackage{here}
\usepackage{psfrag}
\begin{document}

\newcommand{\ket}[1]{|#1 \rangle}

\preprint{APS/123-QED}

\title{Efficient Preparation of Quantum States With Exponential Precision}

\author{Peter Jaksch}
\email{petja@ifm.liu.se}
\affiliation{Department of Physics and Measurement Technology, Link\"oping
University, S-58183 Link\"oping, Sweden.}

\date{\today}

\begin{abstract}
It has been shown that, starting from the state $\ket{0}$, in the
general case, an arbitrary quantum state $\ket{\psi}$ cannot be
prepared with exponential precision in polynomial time. However, we
show that for the important special case when $\ket{\psi}$ represents
discrete values of some real, continuous function $\psi(x)$, efficient
preparation is possible by applying the eigenvalue estimation
algorithm to a Hamiltonian which has $\psi(x)$ as an eigenstate. We
construct the required Hamiltonian explicitly and present an iterative
algorithm for removing unwanted superpositions from the output state
in order to reach $\ket{\psi}$ within exponential accuracy. The method
works under very general conditions and can be used to provide the
quantum simulation algorithm with very accurate and general starting
states.
\end{abstract}


\maketitle

The first step of a quantum algorithm typically consists of resetting 
all registers to $\ket{0}$ followed by the preparation of an initial 
state $\ket{\psi}$. In the general case, it has been shown impossible to 
prepare an arbitrary state with exponential precision in polynomial time 
\cite{kni95}. For special cases, however, this is still possible, 
e.g., when $\ket{\psi}$ represents sampled points of some continuous 
function $\psi(x)$, and certain integrals over $\psi(x)$ can be evaluated 
with exponential precision \cite{zalka}. A useful application 
of such methods is to prepare initial states for quantum simulation 
\cite{zalka}.

In this paper we show that quantum states that represent sampled, 
continuous functions can, under very general conditions, 
be prepared efficiently. We also provide a method for doing this, using
the quantum eigenvalue estimation algorithm by Abrams 
and Lloyd \cite{AL99} and the initial state preparation algorithm by Jaksch 
and Papageorgiou \cite{peter}.

Suppose that $\ket{\psi}$ represents discrete values of some real
function $\psi(x)$. 
The intuitive idea is that if $\ket{\psi}$ is an eigenvector of the 
(discretized) Schr\"odinger equation then it can be obtained by running the 
eigenvalue estimation algorithm with an initial approximation generated by 
the technique in \cite{peter}. The one-dimensional {\it continuous} 
Schr\"odinger equation for
a particle with mass $m$, in a potential $V(x)$, is defined as
\begin{equation}
\left[ -\frac{\hbar^2}{2m} \frac{d^2}{dx^2} + V(x) \right] \phi(x)
= \varepsilon \phi(x), \label{schrodinger}
\end{equation}
where $\varepsilon$ is the energy.
For convenience, we will choose units such that \mbox{$\hbar^2/(2m) = 1$}. 
By defining the potential as 
\begin{equation}
V(x) = \frac{1}{\psi(x)}\frac{d^2\psi(x)}{dx^2}, \label{potential}
\end{equation}
$\psi(x)$ will apparently become an eigenvector with eigenvalue $0$.
Thus, if $0$ is measured from the eigenvalue estimation algorithm we
know that the remaining (unmeasured) qubits will be in the state 
$\ket{\psi}$. More precisely, assume that the continuous
Hamiltonian (the expression in brackets in (\ref{schrodinger})) is 
discretized into a $2^n \times 2^n$ matrix $H$, and that $U = e^{iHt}$
can be implemented {\it exactly} on a quantum computer. 
An arbitrary eigenvector $\ket{\psi_l}$ of $U$ satisfies
\begin{equation}
e^{iHt}\ket{\psi_l} = e^{i\lambda_lt}\ket{\psi_l} = e^{i2\pi \varphi_l} 
\ket{\psi_l}.  
\end{equation}
Hence, $\lambda_lt = 2\pi \varphi_l + n2\pi$, with $n \in Z$
and $\varphi_l \in [0,1)$. We want to prepare the specific state 
$\ket{\psi_0}$ for which $\lambda_0 = 0$ (and $\varphi_0 = 0$).

The discrete Hamiltonian $H$ can be written $H = T + V$, where $T$ is
the second derivative operator, and $V$ is the potential. The norm of
the former is $\|T \| = 2^{2n}$ \cite{demmel}. We make the assumption
that $\|V \| = O(2^{2n})$, so that $\|H \| \le \|T \| + \|V \|\le
2^{2n+p}$ for some integer $p$, independent of $n$. For simplicity,
we also assume that all elements of $V$ are positive
\footnote{It is straightforward to relax this condition to allow for 
negative values of $V$.},
making $H$ positively semidefinite, and that 
$\min_{l \ne 0} \lambda_l = O(1)$, independent of $n$. 
By choosing $t = 1/2^{2n+p}$ all eigenvalues of $U$ will apparently
fall in the first quadrant, i.e., $\varphi_l < 1/4$ for all $l$. 

\begin{figure}[H]
\psfrag{H}{\hspace{-1.5mm}Hd}
\psfrag{in}{$\ket{\Psi_q^s}$}
\psfrag{out}{$\ket{\Psi_{q+1}^s}$}
\psfrag{|0>}{$\ket{0}$}
\psfrag{Uj}{\hspace{-1.3mm}$U^{(s)}$}
\psfrag{0}{$0$}
\centering
\includegraphics[width=7cm]{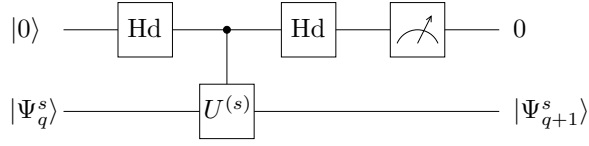}
\caption{Schematic description of one step of the quantum state 
  preparation algorithm. Here, Hd means the Hadamard gate;
  $U^{(s)} = \exp (iH2^st)$. When $q$ reaches some value $k$ it is
  reset to 0 and $s$ is increased with 1.}
\label{prepare}
\end{figure}

Now, suppose that the eigenvalue estimation algorithm is started in
the state 
\begin{equation}
\ket{0} \ket{\Psi_0} = \sum_{l=0}^{2^n-1} d_l \ket{0} \ket{\psi_l},
\end{equation}
where the first register consists of only a single qubit.
At the final step of the algorithm, before measurement, the state 
will be 
\begin{equation}
\sum_{l=0}^{2^n-1} d_l (g(\varphi_l,0)\ket{0} + g(\varphi_l,1)\ket{1}) 
\ket{\psi_l}.
\end{equation}
Here, the function $g$ is defined as (see \cite{peter})
\begin{equation}
g(\varphi_l,j) = \left\{ \begin{array}{ll} \frac{\sin (\pi(2\varphi_l -j))
e^{\pi i(\varphi_l - j/2)}} {2\sin(\pi (\varphi_l-j/2))}, 
& 2\varphi_l \neq j \\
1, & 2\varphi_l = j.
\end{array} \right.
\end{equation}
We remark that for the case $j=0$, $\varphi_l \ne 0$, $g$ reduces to 
$g(\varphi_l,j) = \cos(\pi \varphi_l) e^{i\pi \varphi_l}$.
The probability of measuring the first qubit in the state $\ket{0}$,
ensuring that $\ket{\psi_0}$ is still in the superposition, is
\begin{equation}
P_1 = \sum_{l=0}^{2^n-1} |d_l|^2 |g(\varphi_l,0)|^2.
\end{equation}
Assuming that $0$ is measured, the second register will collapse to
\begin{equation}
\ket{\Psi_1} = \frac{1}{\sqrt{P_1}} \sum_{l=0}^{2^n-1} d_l g(\varphi_l,0) 
\ket{\psi_l}.
\end{equation}
From induction it follows that repeated use of this procedure leads to
\begin{eqnarray}
P_k &=& \frac{1}{P_1 \cdots P_{k-1}} 
\sum_{l=0}^{2^n-1} |d_l|^2 |g(\varphi_l,0)|^{2k} \\
\ket{\Psi_k} &=& \frac{1}{\sqrt{P_1 \cdots P_{k}}}
\sum_{l=0}^{2^n-1} d_l (g(\varphi_l,0))^k \ket{\psi_l}.
\end{eqnarray}
The probability of measuring $0$ every time (probability of success) is
therefore 
\begin{equation}
P_1 \cdots P_{k} = \sum_{l=0}^{2^n-1} |d_l|^2 |g(\varphi_l,0)|^2 
\ge |d_0|^2 |g(\varphi_0,0)|^2 = |d_0|^2.
\end{equation}
The last equality follows since $\varphi_0 = 0$.
With the method for initial state preparation in \cite{peter} we can make
\mbox{$|d_0|^2 > 1/2$}. From the expression for $\ket{\Psi_k}$ it is clear 
that only states with $|g(\varphi_l,0)|$ close to $1$ 
($\varphi_l \le 1/\sqrt{k}$)
will survive as $k$ grows. If we choose, e.g.,
$k = 2n$ all states with $\varphi_l \ge 1/4$ will have their amplitudes
decreased by a factor more than $1/2^n$. By choosing a new value of $t$,
$t' = 2t$ we can reduce a second set of amplitudes exponentially. 
After $2n+p$ such steps we are left with
\begin{eqnarray}
P_k^{(2n+p)} &=& \frac{1}{P_1^{(0)} \cdots P_k^{(0)} \cdots 
P_1^{(2n+p)} \cdots P_{k-1}^{(2n+p)}} \sum_{l=0}^{2^n-1}
|d_l|^2 |g(\varphi_l,0) \cdots g(2^{2n+p}\varphi_l,0)|^{2k} \\
\ket{\Psi_k^{(2n+p)}} &=& \frac{1}{\sqrt{P_1^{(0)} \cdots P_k^{(0)} 
\cdots P_1^{(2n+p)} \cdots P_{k}^{(2n+p)}}}
\sum_{l=0}^{2^n-1} d_l (g(\varphi_l,0) \cdots g(2^{2n+p}\varphi_l,0))^k 
\ket{\psi_l} \\ 
P(\mathrm{success}) &=& \sum_{l=0}^{2^n-1}
|d_l|^2 |g(\varphi_l,0) \cdots g(2^{2n+p}\varphi_l,0)|^{2k} 
\ge |d_0|^2 |g(\varphi_0,0) \cdots g(2^{2n+p}\varphi_0,0)|^{2k} = |d_0|^2.
\end{eqnarray}
The condition that $\min_{l \ne 0} \lambda_l = O(1)$ ensures that at least 
one of the terms $g(2^j\varphi_l,0)$ will satisfy $g(2^j\varphi_l,0)\le 1/A$,
for some $A>1$, independent of $n$ and $k$. 
Hence, if $k$ is polynomial in $n$ the only term
in the superposition that is not exponentially reduced is $\ket{\psi_0}$.
The algorithm can be implemented as a sequence of $(2n+p)k$ circuits of
the type in \mbox{Fig. \ref{prepare}}, where $U^{(s)} = \exp (iH2^st)$.

The requirement that $U$ can be implemented exactly can be
weakened to requiring only exponential precision. In Fig. \ref{prepare} 
the state before measurement is 
\begin{equation}
\frac{1}{2}\ket{0}(I+U^{(s)})\ket{\Psi_q^{(s)}}
+ \frac{1}{2}\ket{1}(I-U^{(s)})\ket{\Psi_q^{(s)}}.
\end{equation} 
The probability of measuring $0$ is $P_q^{(s)}=\| I+U^{(s)} \|^2/4$.
For this outcome the second register will collapse to
\begin{equation}
\frac{1}{2\sqrt{P_q^{(s)}}}(I+U^{(s)})\ket{\Psi_q^{(s)}}.
\end{equation} 
Suppose now that $\tilde{U}^{(s)} = U^{(s)} + \Delta$ can be implemented
with exponential precision, i.e., $\| \Delta \| = O(1/2^n)$. 
For this case, we derive a lower bound on the
probability of success. First, we make the following observation:
\begin{eqnarray}
\left| \sqrt{\tilde{P}_q^{(s)}} - \sqrt{P_q^{(s)}} \right| &=& \frac{1}{2}
\left| \|(I+\tilde{U}^{(s)})\ket{\Psi_q^{(s)}}\|
- \|(I+U^{(s)})\ket{\Psi_q^{(s)}}\| \right| \nonumber \\
&\le& \frac{1}{2} \|(U^{(s)}-\tilde{U}^{(s)})\ket{\Psi_q^{(s)}}\|
= \frac{1}{2} \| \Delta \ket{\Psi_q^{(s)}}\| \le \frac{\|\Delta \|}{2}. 
\label{ctilde}
\end{eqnarray}
From (\ref{ctilde}) it follows directly that $| \tilde{P}_q^{(s)} - P_q^{(s)}| < \|\Delta \|$.
Using this result we get 
\begin{equation}
\tilde{P}_1^{(0)} \cdots \tilde{P}_{k}^{(2n+p)} 
\ge (P_1^{(0)} - \|\Delta \|) \cdots (P_k^{(2n+p)}-\|\Delta \|)
= P_1^{(0)} \cdots P_{k}^{(2n+p)} (1 - |O(\|\Delta \|)|)^{k(2n+p)},
\end{equation}
which is bounded by a constant for exponentially small $\|\Delta \|$
\footnote{In fact, this is true even for polynomially small $\|\Delta \|$}.
Using (\ref{ctilde}) and the definition of $P_q^{(s)}$ we may also derive 
an upper bound on the difference 
between $\ket{\tilde{\Psi}_{q+1}^{(s)}}$ and $\ket{\Psi_{q+1}^{(s)}}$
\begin{eqnarray}
\| \ket{\tilde{\Psi}_{q+1}^{(s)}} - \ket{\Psi_{q+1}^{(s)}}\| &=& \frac{1}{2}
\left\| \frac{1}{\sqrt{\tilde{P}_q^{(s)}}}
(I + U^{(s)} + \Delta)\ket{\Psi_q^{(s)}}
- \frac{1}{\sqrt{P_q^{(s)}}}(I+U^{(s)})\ket{\Psi_q^{(s)}} \right\|
\nonumber \\
&\le& \frac{\left| \sqrt{\tilde{P}_q^{(s)}} - \sqrt{P_q^{(s)}} \right| }
{\sqrt{ \tilde{P}_q^{(s)} }} + \frac{1}{2\sqrt{\tilde{P}_q^{(s)}}}
\| \Delta \| \le \frac{\| \Delta \|}{\sqrt{ \tilde{P}_q^{(s)} }} = 
O(\| \Delta \|). \label{psitilde}
\end{eqnarray}
It can be shown \cite{bv97} 
that the total error of the algorithm is at most the sum of the errors
of the individual blocks. Hence, with the previous choice of $k$,
\begin{equation}
\| \Delta_{\mathrm{tot}} \| \le (2n+p)k O(\| \Delta \|) = O(n^2/2^n).
\end{equation}

We now consider the implementation of $U$. Suppose that we use an m:th 
order splitting formula (see \cite{dragt}) to approximate $U^{(s)}$:
\begin{equation}
U^{(s)} = e^{iH2^st} = \prod_{j=1}^m e^{iw_jG_j2^st} - \Delta,
\label{splitting}
\end{equation}
where $w_j$ are weight factors and $G_j$ is either $T$ or $V$.
Defining the product on the right hand side $\tilde{U}^{(s)}$
it follows from the definition of the matrix exponential that $\Delta$
is the Taylor expansion of $\tilde{U}^{(s)} - U^{(s)}$ beyond m:th order. 
Provided that $V(x)$ can be efficiently computed to exponential precision 
on a classical computer, both $e^{iVt}$ and $e^{iTt}$ can be efficiently
calculated on a quantum computer, as described in \cite{zalka}.

The norm of $\Delta$ is of order $O((2^st)^m (\| T \| + \| V \| )^m)$.
For some $B_* \ge 1$ we impose the (not very restrictive) condition
$\|T \|_* + \|V \|_* \le B_* \lambda_*$, where the norm is defined on
a subspace spanned by a number of basis vectors $\ket{\psi_l}$ of $H$,
containing not only the ground state, and $\lambda_*$ is the maximum
eigenvalue of these vectors
\footnote{The inequality $\|T \| + \|V \| \le \|H \| + 2\|V \|$ is
trivial.  Suppose that $\psi$ is some eigenvector of $H$ with
eigenvalue $\lambda$. For the case $V > \lambda$, $\psi$ decays
exponentially as $e^{\lambda-V}$. Hence, if $\lambda$ is the maximum
eigenvalue on some subspace spanned by eigenvectors of $H$, the norm
$\|V \|$ will normally be of order $\lambda$. The right hand side of
the inequality will then also be of order $\lambda$ on this
subspace.}.  Defining $B = \sup B_*$, where the supremum is taken over
all subspaces, we choose $t$ as $t = 1/(B2^{n+p})$. For $k = -\lceil
n/\log_2(\pi/(4B)) \rceil$ all states with $\varphi_l > 1/(4B)$ will
have their amplitudes decreased by a factor more than $1/2^n$. 

We are now in position to modify the previous error analysis. 
If $2^stB\lambda_* \le 1/C$, for some $C>1$ independent of $m$,
the error in (\ref{splitting}) will be exponentially small, of order 
$O(1/C^m)$, on the subspace where $\lambda_*$ is defined.  
\begin{eqnarray}
\left| \sqrt{\tilde{P}_q^{(s)}} - \sqrt{P_q^{(s)}} \right| 
&\le& \frac{1}{2} \|\Delta \ket{\Psi_q^{(s)}} \| \le \frac{1}{2|d_0|}  
\left\| \Delta \sum_{l=0}^{2^n-1} d_l (g(\varphi_l,0) \cdots 
g(2^{s}\varphi_l,0))^q \ket{\psi_l} \right\| \nonumber \\ &\le&
O\left( \frac{1}{C^m} \right) + \frac{\sqrt{2}}{2} \| \Delta \|
\left\| \sum_{2^stB\lambda_l \ge 1} d_l (g(\varphi_l,0) \cdots 
g(2^{s}\varphi_l,0))^k \ket{\psi_l} \right\|. \label{approx}
\end{eqnarray}
For the eigenvectors in the sum in (\ref{approx}), if $2^st\lambda_l
\ge 2\pi$, then at least one of the arguments $2^j \varphi_l$ will be
$O(1)$, and, hence, the amplitude of this state will be exponentially
reduced. If $2^st\lambda_l < 2\pi$, on the other hand, then $2^s
\varphi_l \ge 1/(2\pi B)$. We make the assumption that $B$ is bounded from 
above by a polynomial in $n$. Hence,
\begin{equation}
\left| \sqrt{\tilde{P}_q^{(s)}} - \sqrt{P_q^{(s)}} \right| = 
O\left( \frac{1}{C^m} \right) + O(1)O\left( \frac{1}{D^k} \right), 
\label{final}
\end{equation}
where $D$ is greater than $1$ independent of $k$. The fact that $\|
\Delta \| = O(1)$ follows since $\| U^{(s)} \| = \| \tilde{U}^{(s)} \|
= 1$. By choosing $k$ as above, and $m$ polynomial in $n$ the right
hand side of (\ref{final}) can be made exponentially small in $n$. In
the same way it is possible to prove that $\| \ket{\tilde{\Psi}_k^{(s)}} -
\ket{\Psi_k^{(s)}} \|$ will also be exponentially small in $n$, and,
thus, the same holds for the total error. In practice, one would have to
guess the value of $B$. However, as long as the guessed value is larger than
the true $B$ it does not have to be very accurate.

For the sake of clarity, we have not yet discussed the issue of 
boundary conditions, but rather assumed implicitly that $\psi(0) = 
\psi(1) = 0$. This will be the case for most problems of practical 
interest and is also a requirement for the method in \cite{peter} to work.
Here, however, we will briefly outline a scheme for generalizing our method 
to functions with non-zero boundary conditions. One way of doing this is
to extend the function to be prepared by smoothly attaching exponentially 
decaying tails on both sides. Practically, this can be achieved by 
adding, say, two extra qubits to the register holding the function to
be prepared. As an example, states starting with qubits 00 will 
represent the left hand side tail; states starting with qubits 01 will
represent the original function $\psi$; and states starting with qubits 10
and 11 will represent the right hand side tail. At the end of the preparation
we will be left with the extended function. To get rid of the tails we can
now measure the two qubits that we added to the register. With high 
probability this measurement will yield 01, implying that the extended 
state has collapsed to the original function $\psi$.

I am grateful to Anargyros Papageorgiou and Jan-{\AA}ke Larsson for
very fruitful discussions and helpful comments.


\end{document}